\RequirePackage{fix-cm}
\documentclass[smallextended]{svjour3}       
\smartqed  
\usepackage{graphicx}
\begin{document}

\title{Quantum gravity effect on the Hawking radiation of charged rotating BTZ black hole}


\author{Ganim Gecim \and Yusuf Sucu}


\institute{G. Gecim \at
              Department of Physics, Faculty of Science, Akdeniz
University, 07058 Antalya, Turkey \\
              \email{gecimganim@gmail.com}
           \and
           Y. Sucu \at
              Department of Physics, Faculty of Science, Akdeniz
University, 07058 Antalya, Turkey \\
              \email{ysucu@akdeniz.edu.tr}
}


\maketitle

\begin{abstract}
In this study, the quantum gravity effect on the tunnelling radiation of
charged massive spin-0 scalar particle from 2+1 dimensional charged rotating Banados-Teitelboim-Zanelli (BTZ) black hole is
looked into by using the Hamilton-Jacobi approach. For this, we calculate
the modified Hawking temperature of the black hole by using the modified
Klein-Gordon equation based on the Generalized Uncertainty Principle (GUP), and we noticed that the modified Hawking
temperature of the black hole depends not only on the black hole properties,
but also on the angular momentum, energy, charge and mass of
the tunnelling scalar particle. Using the modified Hawking temperature, we discussed the stability of the black
hole in the context of the modified heat capacity, and observed that it might
undergo both first and second-type phase transitions in the presence of the quantum
gravity effect, but just a first-type transition in the absence of the quantum
gravity effect. Furthermore, we investigated the modified Hawking temperature of the black hole by using the tunnelling processes of the charged massive Dirac and vector boson particles. We observed that scalar, Dirac and vector particles are tunnelled from the black hole completely differently from each other in the presence of the quantum gravity effect.
\end{abstract}

\section{Introduction}\label{intro}
It is one of the most important problems of modern physics to construct
a self-consistent quantum gravity theory by merging quantum mechanics with
general relativity. With such a theory, we expect to clarify the fundamental
physical problems in gravity, e.g. the origin of the universe and the final
stage of a black hole. At present, there are several theories as candidates
such as the string theory and the loop quantum gravity theory which exhibit some
features already expected from a self-consistent quantum gravity theory. The
common feature of these theories is that they all point out the existence of
a minimum observable length in the order of the Planck scale \cite{1,2,3,4,5,6,7,8,b1,b2}. The existence of such a minimal length leads to the
generalized Heisenberg uncertainty principle (GUP), and it causes in some
modifications on the quantum mechanical relations \cite{6,7}. Together with
the modifications, the intrinsic properties of a particle as an extended
object begin to emerge in the quantum gravity effects \cite{33,34,35,49,50,51,52}.

With the well-known studies of Bekenstein, Bardeen, Carter and Hawking, a
black hole has been considered as a thermodynamical system \cite{9,10,11,12,13,14,15}. From these studies, in particular, in the Hawking's
studies that consider quantum mechanical methods in a curved spacetime it
was proved that the thermal radiation of a black hole, known as Hawking
radiation, stem from the quantum vacuum fluctuation near the black hole
horizon. Since then, to get this radiation it has been put forward various
alternative approaches. For instance, the Hamilton-Jacobi approach based on
the tunnelling process of an elementary particle throughout the classically
forbidden trajectory from inside to outside of a black hole horizon is one
of the effective ways to derive the radiation \cite{16,17,18,19,20,21,22,23,24,25,26,27,28,29,30,a1,a2,a3,a4,a5,a6,a7}. In all the
studies realized by the context of the standard Heisenberg uncertainty
principle, it has been seen that the Hawking radiation of a black hole
depends on only the properties of black hole. However, in the studies
performed in the GUP context, it has been come out that the Hawking
radiation is depended on both the properties of the black hole and the
tunneling particle \cite{33,34,35,49,31,32,36,37,38,39,40,41,42,43,44,45,46,47,48}. At that case, it
can be determined that what kind of particle tunnels from a black hole \cite{50,51}. With this motivation, in the study, we will investigate the Hawking
radiation of the $2+1$-dimensional charged rotating
Banados-Teitelboim-Zanelli (hereafter CR-BTZ) black hole by using the
quantum tunnelling process of the charged massive scalar, and subsequently, from the derived results we will discuss the stability of the black hole under the GUP effects. Furthermore, in the Appendix A and B, Hawking radiation of the CR-BTZ black hole will be derived by means of the tunneling process of the charged massive Dirac and vector boson particles under the GUP effects, respectively, and then, the results will be compared with that of the scalar particle.

The structure of this paper is as follows: In section-2, in the context of the GUP, we
find the modified Hawking temperature of the black hole by using the
tunnelling process of the charged massive scalar particle (Also, the quantum gravity effect on the tunnelling of the Dirac and vector boson particles is discussed in the Appendix A and B, respectively.). In section-3, using the modified Hawking temperature of the charged massive scalar particle, we calculate the modified heat
capacity of the black hole and, subsequently, discuss the local stability of the black
hole. In section-4, we summarize the results.

\section{Tunneling of the charged massive scalar particle from the CR-BTZ black hole}\label{Scalar}

The BTZ black hole is an exact solution of the $(2+1)$-dimensional
Einstein-Maxwell gravity theory and its charged and rotating case includes
more rich mathematical and physical structure than its non-charged and
non-rotating cases \cite{BTZ1,BTZ2,clement1,clement2}. Discovery of BTZ black hole solutions represents one of the cornerstones for research on $(2+1)$ dimensional gravitation theories. Beside Einstein-linear Maxwell theory, Einstein-nonlinear electrodynamics theory admits the BTZ black hole as a solution, also \cite{Hendi1,Hendi2}. The nonlinear electrodynamics theory was firstly introduced by Born and Infeld to obtain a finite value for the self-energy and radius of the electron \cite{Born}. The coupling nonlinear electrodynamics theory to Einstein gravitation theory was firstly worked out by Hoffmann \cite{Hoffmann}. Furthermore, after Bardeen firstly introduced a regular black hole solution in the context of Einstein-nonlinear electrodynamics theory \cite{Bardeen}, the interest on this theory has increased and new black hole solutions found in 3+1 dimensional as well as 2+1 dimensional \cite{Beato1,Myung,Habib1,Habib2,He,Sajadi}.

In order to emerge quantum gravity effect on the Hawking temperature of the CR-BTZ black hole, we start by writing the explicit form of its spacetime background as follows \cite{BTZ1};
\begin{eqnarray}
ds^{2}=f(r)dt^{2}-\frac{1}{f(r)}dr^{2}-r^{2}\left[d\varphi+N^{\varphi}(r)dt\right]^{2}, \label{BTZBH1}
\end{eqnarray}
where the $f(r)$ and $N^{\varphi}(r)$ are
\begin{eqnarray*}
f(r)&=&-M+\frac{r^{2}}{l^{2}}+\frac{J^{2}}{4r^{2}}-\frac{Q^{2}}{2}\ln(\frac{r}{l}), \\ \ \ \ \
N^{\varphi}(r)&=&-\frac{J}{2r^{2}},
\end{eqnarray*}
where $l$ is radius related to the cosmological constant as $l^{2}$=$%
-1/\Lambda $, and, $M$, $Q$ and $J$ are the mass, electric charge and
angular momentum of the black hole, respectively. The mass and the angular velocity of the outer horizon of the black hole are
given as follows;
\begin{eqnarray}
M&=&\frac{r_{+}^{2}}{l^{2}}+\frac{J^{2}}{4r_{+}^{2}}-\frac{Q^{2}}{2}\ln(\frac{r_{+}}{l}),\\ \ \ \ \ \label{mass}
\Omega_{+}&=&\frac{J}{2r_{+}^{2}}, \label{angularV}
\end{eqnarray}
respectively, where $r_{+}$ is the radius of the outer horizon.

As point out in \cite{23,Jiang1,Jiang2,Chen}, the spacetime and matter near the outer horizon of the stationary black hole are dragged by the hole. Also, in the stationary black hole, the outer horizon does not coincide with the infinite redshift surface. These difficulties can be avoided by implementing a dragging coordinate transformation \cite{Jiang1,Jiang2,Chen,Rizwan,Mahanta,Jan,Sakalli,Anacleto,Li}. Hence, we carry out a dragging coordinate transformation $d\phi=d\varphi+N^{\varphi}dt$ by
means of the Killing vectors, $\left(\partial_{t}\right)$ and $\left(\partial_{\varphi}\right)$. Then, the metric takes the following form
\begin{equation}
ds^{2}=f(r)dt^{2}-\frac{1}{f(r)}dr^{2}-r^{2}d\phi^{2}.\label{BTZBH2}
\end{equation}

In order to investigate the quantum gravity effects on the tunneling process
of the scalar particles from the black hole by using the GUP relations,
the standard Klein-Gordon equation is modified as
\begin{eqnarray}
\left[\left(-i\hbar\right)^{2}\partial_{i}\partial^{i}-M_{0}^{2}-iq\hbar(\partial_{\mu}A^{\mu})-2 iq\hbar A^{\mu}\partial_{\mu}+q^{2}A_{\mu}A^{\mu}\right] \left[\widetilde{\Phi}-2\alpha\left(-\hbar^{2}\partial
_{i}\partial^{i}+M_{0}^{2}\right)\widetilde{\Phi}\right]\nonumber \\+\left(i\hbar\right)^{2}\partial_{t}\partial^{t}\widetilde{\Phi}=0, \label{MKG1}
\end{eqnarray}
where $\widetilde{\Phi }$, $M_{0}$ and $A_{\mu }$ are the modified wave
function, mass of the scalar particle and the electromagnetic potential
vector, respectively. In addition, $\alpha=\alpha_{0}/M_{p}^{2}$ with the $M_{p}^{2}$ and $\alpha_{0}$ are the Planck mass and dimensionless
parameter, respectively \cite{6,7}. For the non-vanishing component of the
electromagnetic potential, $A_{0}$, and $\alpha ^{2}=0,$ the explicit form
of the modified Klein-Gordon equation can be simplified by the following
way:
\begin{eqnarray}
\hbar^{2}\partial_{t}\partial^{t}\widetilde{\Phi}+ \hbar^{2}\partial_{i}\partial^{i}\widetilde{\Phi}+ 2\alpha \hbar^{4}\partial_{i}\partial^{i}(\partial_{i}\partial^{i}\widetilde{\Phi})-(q^{2}A_{0}A^{0}-M_{0}^2)(1-2\alpha M_{0}^2)\widetilde{\Phi}\nonumber \\+2iq\hbar A^{0}(1-2\alpha M_{0}^2)\partial_{0}\widetilde{\Phi}+4iq\alpha\hbar^{3}A^{0}\partial_{0}(\partial_{i}\partial^{i}\widetilde{\Phi})-2\alpha q^{2}\hbar^{2}A_{0}A^{0}\partial_{i}\partial^{i}\widetilde{\Phi}=0, \label{MKG01}
\end{eqnarray}
Then, the modified Klein-Gordon equation in the CR-BTZ black hole background
and the electromagnetic potential, $A_{0}$=$-Q\ln(\frac{r}{l}),$ is written as
\begin{eqnarray}
\frac{\hbar^{2}}{f}\frac{\partial^{2}\widetilde{\Phi }}{\partial t^{2}}-\frac{\hbar^{2}}{r^{2}}
\frac{\partial^{2}\widetilde{\Phi}}{\partial\phi^{2}}+2\alpha\hbar^{4}f\frac{\partial^{2}}{\partial r^{2}}\left(f\frac{\partial^{2}\widetilde{\Phi}}{\partial r^{2}}\right)
+\frac{2\alpha\hbar^{4}}{r^{2}}\frac{\partial^{2}}{\partial\phi^{2}}\left(\frac{1}{r}\frac{\partial^{2}
\widetilde{\Phi}}{\partial\phi^{2}}\right) \nonumber \\-\hbar^{2}f\frac{\partial^{2}\widetilde{\Phi}}{\partial r^{2}}-\left(\frac{q^{2}A_{0}^{2}}{f}-M_{0}^{2}\right)\left(1-2\alpha M_{0}^{2}\right) \widetilde{\Phi}+2iq\hbar\frac{A_{0}}{f}\left(1-2\alpha M_{0}^{2}\right)\frac{\partial\widetilde{\Phi}}{\partial t} \nonumber \\-4iq\alpha\hbar^{3}\frac{A_{0}}{f}\frac{\partial}{\partial t}\left(f\frac{\partial^{2}\widetilde{\Phi }}{\partial r^{2}}+\frac{1}{r}\frac{\partial^{2}\widetilde{\Phi}}{\partial\phi^{2}}\right)+2\alpha q^{2}\hbar^{2}\frac{A_{0}^{2}}{f}\left(f\frac{\partial^{2}\widetilde{\Phi}}{\partial r^{2}}+\frac{1}{r}\frac{\partial^{2}\widetilde{\Phi}}{\partial\phi^{2}}\right)=0.
\label{ScalarD2}
\end{eqnarray}
After that, if the modified wave function of the scalar particle, $\widetilde{\Phi}\left(t,r,\phi \right)$, is defined as
\begin{eqnarray}
\widetilde{\Phi}\left(t,r,\phi\right)=A\exp\left(\frac{i}{\hbar}S\left(t,r,\phi\right)\right), \label{ansatz}
\end{eqnarray}
where $A$ is a constant and $S(t,r,\phi )$ is the classical action function.
Substituting the wave function expression into the Eq.(\ref{ScalarD2}) and
subsequently neglecting the higher order terms of $\hbar $, we get the
modified Hamilton-Jacobi equation as follows;
\begin{eqnarray}
\left(\frac{\partial S}{\partial t}\right)^{2}-f^{2}\left(\frac{\partial S}{\partial r}\right)^{2}-\frac{f}{r^{2}}\left(\frac{\partial S}{\partial
\phi}\right)^{2}+\left(q^{2}A_{0}^{2}-M_{0}^{2}f\right)-\alpha \frac{2f}{r^{4}}\left(\frac{\partial S}{\partial \phi}\right)^{4} \nonumber
\\ +2qA_{0}\left(1-2\alpha M_{0}^{2}\right)\left(\frac{\partial S}{\partial t}\right)-\alpha \left[2M_{0}^{2}\left(q^{2}A_{0}^{2}-M_{0}^{2}f\right)+2f^{3}\left(\frac{\partial S}{\partial r}\right) ^{4}\right]\nonumber
\\ +\alpha\left[2fq^{2}A_{0}^{2}\left(\frac{\partial S}{\partial r}\right)^{2}+4fqA_{0}\left(\frac{\partial S}{\partial r}\right)^{2}\left(\frac{\partial S}{\partial t}\right)+\frac{2q^{2}A_{0}^{2}}{r^{2}}\left(\frac{\partial S}{\partial \phi}\right)^{2}\right]\nonumber
\\+\alpha \frac{4qA_{0}}{r^{2}}\left(\frac{\partial S}{\partial \phi}\right)^{2}\left(\frac{\partial S}{\partial t}\right)=0
\label{HamiltonM}
\end{eqnarray}
To solve this equation, by using separation of variable method, the $S\left(
t,r,\phi \right) $ can be separated as $S\left( t,r,\phi \right) $=$%
-(E-j\Omega _{+})t+j\phi +W(r)+C$ in which $C$ is a complex constant, $%
\Omega _{+}$ is the angular velocity of the outer horizon, and $E$, $j$ and $%
W(r)$=$W_{0}(r)+\beta W_{1}(r)$ are energy, angular momentum and radial
trajectory of the particle, respectively \cite{50,51,39}. After some
calculations, the radial trajectory $W_{\pm }(r)$ are obtained as
\begin{equation}
W_{\pm }(r)=\pm \int \frac{\sqrt{(E-j\Omega_{+}-qA_{0})^{2}-f\left(M_{0}^{2}+j^{2}/r^{2}\right)}}{f}\left[ 1+\alpha \Theta\right]dr,
\label{MEqq1}
\end{equation}
where $W_{+}(r)$ and $W_{-}(r)$ correspond to the outgoing and incoming
particle trajectories, respectively, and the abbreviation $\Theta $ is
\begin{eqnarray*}
\Theta&=&\frac{f^{2}\left(M_{0}^{4}-j^{4}/r^{4}\right)+qA_{0}f(M_{0}^{2}-j^{2}/r^{2})\left[2(E-j\Omega_{+})-qA_{0}\right]}{f\left[(E-j\Omega_{+}-qA_{0})^{2}-f\left(M_{0}^{2}+j^{2}/r^{2}\right) \right]}\\
&&-\frac{\left[(E-j\Omega_{+})^{2}-f\left(M_{0}^{2}+j^{2}/r^{2}\right) \right]}{f}.
\end{eqnarray*}
As $f(r)\approx (r-r_{+})f^{\prime }(r_{+})$ near the outer horizon, the $W_{\pm}(r)$ are computed as
\begin{equation}
W_{\pm }(r_{+})=\pm i\pi \frac{\left[E-j\Omega_{+}+qQ\ln(\frac{r_{+}}{l})\right]}{H}\left[ 1+\alpha\Sigma\right],\label{MEqq2}
\end{equation}
with the abbreviations $\Sigma$ and $H$ are
\begin{eqnarray*}
\Sigma&=&\frac{4j^{2}qQH\ln(\frac{r_{+}}{l})\left[2(E-j\Omega_{+})+qQ\ln(\frac{r_{+}}{l})\right]+3H(E-j\Omega_{+})^{2}r_{+}^{2}\left[M_{0}^{2}+j^{2}/r_{+}^{2}\right]}{2Hr_{+}^{2}\left[E-j\Omega_{+}+qQ\ln(\frac{r_{+}}{l})\right]^{2}}\\
&&-\frac{qQ(E-j\Omega_{+})^{2}}{Hr_{+}\left[E-j\Omega_{+}+qQ\ln(\frac{r_{+}}{l})\right]},\\
&& \\ \ \
H&=&\frac{2r_{+}}{l^{2}}-\frac{J^{2}}{2r_{+}^{3}}-\frac{Q^{2}}{2r_{+}},
\end{eqnarray*}
respectively. On the other hand, the tunneling probabilities of particles
crossing the outer horizon are given by
\begin{eqnarray}
P_{out}=\exp \left[-\frac{2}{\hbar}ImW_{+}(r_{+})\right],
\nonumber \\
P_{in}=\exp \left[-\frac{2}{\hbar}ImW_{-}(r_{+})\right]. \label{Equation16}
\end{eqnarray}
Hence, the tunneling probability of the particle is
\begin{equation}
\Gamma=e^{-\frac{2}{\hbar }ImS}=\frac{P_{out}}{P_{in}}=e^{-\frac{E_{total}}{T^{KG}_{H}}},\label{MEqq11}
\end{equation}
where $E_{total}$ is total energy of the scalar particle, and $T_{H}^{^{KG}}$
is the modified Hawking temperature of the outer horizon for the scalar
particle \cite{53,54,55,56}. Then, the tunneling probability of the charged
massive scalar particle for the black hole is written as
\begin{equation}
\Gamma =\exp \left(-\frac{4\pi [E-j\Omega_{+}+qQ\ln(\frac{r_{+}}{l})]}{\hbar H}\left[ 1+\alpha \Sigma\right]\right)\label{MEqq3}
\end{equation}
and thus the modified Hawking temperature, $T_{H}^{KG}$, is obtained as
follows
\begin{equation}
T_{H}^{KG}=\frac{T_{H}}{\left[1+\alpha \Sigma\right]}.\label{MEqq4}
\end{equation}
where $T_{H}$ is the standard Hawking temperature of the black hole and its explicit expression is
\begin{eqnarray}
T_{H}=\frac{\hbar}{4\pi}\left(\frac{2r_{+}}{l^{2}}-\frac{J^{2}}{2r_{+}^{3}}-\frac{Q^{2}}{2r_{+}}\right). \label{std}
\end{eqnarray}
Furthermore, neglecting the higher order $\alpha $ terms (since $\alpha \ll 1$), we find the modified Hawking temperature of the black hole as follows;
\begin{equation}
T_{H}^{KG}\simeq T_{H}\left[1-\alpha \Sigma\right]. \label{MEqq5}
\end{equation}
This result indicates that the modified Hawking temperature of the charged
massive scalar particle is lower than the standard Hawking temperature.
Moreover, it shows that the modified Hawking temperature depends on not only
the black hole but also the tunneling particle properties. It is important to notice that the modified Hawking temperature of the charged
massive scalar particle is different from that of the both Dirac and vector boson particles (see Eqs.(\ref{MEqq13}) and (\ref{MT}) in the Appendix A and B, respectively.).

\section{Stability analysis}\label{stab}

The local stability of a black hole can be analyzed by the heat capacity
\cite{Cai}. If the heat capacity is positive, then the black hole is locally
stable or else it is unstable \cite{Dehghani,Hendi,Miao}. For this reason,
to discuss the stability of the CR-BTZ black hole, we firstly calculate its
modified heat capacity.

The heat capacity at constant charge, $Q$, and
constant angular momentum, $J$, of the black hole can be expressed by using
the following relation
\begin{eqnarray}
C_{Q,J}=\left(\frac{\partial M}{\partial T_{H}}\right)_{Q,J} \label{capacity}
\end{eqnarray}
where $M$ and $T_{H}$ are the mass and the standard Hawking temperature of
the black hole, respectively. Making a comparison to the quantum gravity effects with
respect to the standard ones, we derive the modified heat capacity of the
black hole from the modified Hawking temperature of the charged massive
scalar particle. Also, it is important to note that the heat capacity derived from
the modified Hawking temperature is going to be different to each particle.
Using Eqs.(\ref{mass}) and (\ref{MEqq4}), the modified heat capacity of the
black hole ($C_{Q,J}^{^{\prime }}$) becomes
\begin{eqnarray}
C_{Q,J}^{'}=\frac{2\pi}{r_{+}\hbar\left[E-j\Omega_{+}+qQ\ln(\frac{r_{+}}{l})\right]} \frac{\left(\mathcal{X}-\mathcal{Y}\right)^{2}}{\left(\mathcal{A}-\mathcal{B}\right)}. \label{capacity1}
\end{eqnarray}
where the constant $\mathcal{X}$, $\mathcal{Y}$, $\mathcal{A}$ and $\mathcal{B}$ are functions of the $E,J,j,l,q,Q,M_{0},r_{+}$ and $\alpha$ (see
Appendix C).

According to Eq.(\ref{capacity1}), the CR-BTZ black hole may undergo both the first and second-type phase transition in the presence of the quantum gravity effect.
The modified heat capacity vanishes for $\mathcal{X}=\mathcal{Y}$ and $\mathcal{A}\neq\mathcal{B}$. If, $\mathcal{A}<\mathcal{B}$, the modified heat capacity is negative. However, If, $\mathcal{A}>\mathcal{B}$, then modified heat capacity is positive. Hence, the black hole undergoes a first-type phase transition to become stabile. On the other hand, the modified heat capacity diverges at point $\mathcal{A}=\mathcal{B}$ for $\mathcal{X}\neq\mathcal{Y}$. Then, in the presence of the quantum
gravity effect, we can mention that the black
hole has a second-type phase transition, also. To elucidate the stability properties of the black hole, if
we plot the $C_{Q,J}^{^{\prime}}$ heat capacity for the special values of
the constants $E, J, j, l, q, Q, M_{0}$ and $\alpha$ since its expression is
very complicated, then we can see that the black hole undergoes both first-type (Fig.1) and a
second-type phase transitions in the presence of the quantum gravity effect
(solid line in Fig. 2).
\begin{figure}[htp!]
\centering
\includegraphics[width=0.55\textwidth]{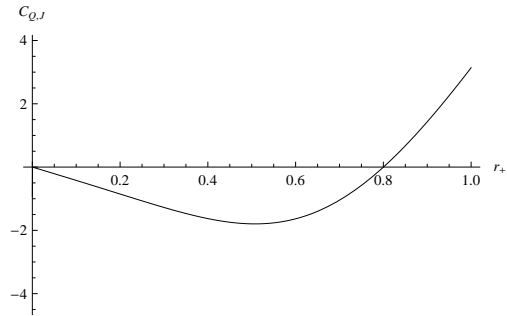}
\caption{$C_{Q,J}$-$r_{+}$ curve. The modified heat capacities of the CR-BTZ black hole. We set $E=6, M_{0}=0.2,q=0.3,\alpha=10^{-9},Q=l=j=J=\hbar=1$.}
\label{Figure-1}
\end{figure}

\begin{figure}[htp!]
\centering
\includegraphics[width=0.55\textwidth]{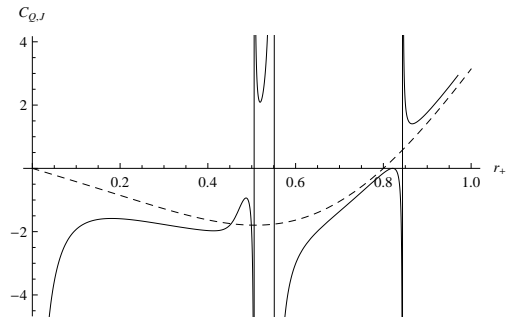}
\caption{$C_{Q,J}$-$r_{+}$ curve. The dashed and solid lines are correspond to the standard and the modified heat capacities of the CR-BTZ black hole, respectively. We set $E=4, M_{0}=2,q=3, \alpha=10^{-2},Q=l=j=J=\hbar=1$.}
\label{Figure-2}
\end{figure}

On the other hand, in the absence of the GUP effect, i.e. $\alpha =0$, the
modified heat capacity of the black hole reduces to the standard one \cite{akbar}:
\begin{eqnarray}
C_{Q,J}=\frac{4\pi r_{+}}{\hbar}\frac{\left(4r_{+}^4-J^2l^2-Q^2l^2r_{+}^2\right)}{\left(4r_{+}^4+3J^2l^2+Q^2l^2r_{+}^2\right)}. \label{capacity2}
\end{eqnarray}
Then, the standard heat capacity vanishes at the point $4r_{+}^{4}=J^{2}l^{2}+Q^{2}l^{2}r_{+}^{2}$. As depicted by the dashed lines in both
Fig. 2 and 3, the black hole is unstable in the
region $0<r_{+}<0.8$, while it is stable in the region $0.8<r_{+}$. Therefore, we can say that the black hole undergoes only the first type
of phase transition in order to become stable in the absence of the GUP
effect.

However, in the presence of the quantum gravity effect, we see that there
are two regions, $0<r_{+}<0.51$ and $0.55<r_{+}<0.84$, in which the black
hole is unstable and that the black hole is stable in the rest two regions
that are $0.51<r_{+}<0.55$ and $0.84<r_{+}$. On the other hand, it is
important to emphasize that the black hole is stable in the region, $0.51<r_{+}<0.55,$ due to quantum gravity effect while it is unstable in this
region in the absence of quantum gravity effect.

\begin{figure}[htp!]
\centering
\includegraphics[width=0.55\textwidth]{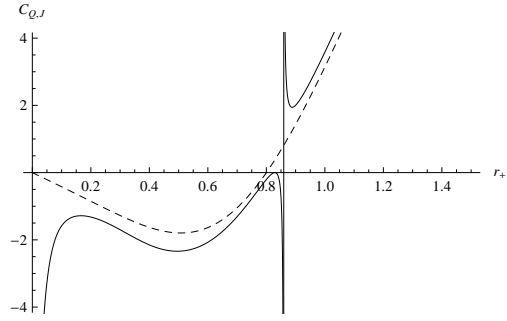}
\caption{$C_{Q,J}$-$r_{+}$ curve. The dashed and solid lines are correspond to the standard and the modified heat capacities of the CR-BTZ black hole, respectively. We set $E=4, M_{0}=2,q=3,\alpha=10^{-2},Q=l=j=\hbar=1,J=-1$.}
\label{Figure-3}
\end{figure}

Also, in the presence of the quantum gravity effect, if the angular momentum
of the black hole is negative, i.e. $J<0$, the black hole becomes unstable
in the region $0.51<r_{+}<0.55$ (solid line in Fig. 3), but, in the
absence of the quantum gravity effect, the stability properties of the black
hole doesn't change, i.e, the black hole is unstable in the region $0<r_{+}<0.84$ and stable in the region $0.84<r_{+}$ (dashed line in
Fig. 3). As is delineated in Fig. 2 and 3, under the GUP effects, it has
been seen that the black hole at $r_{+}=0.84$ undergoes second-type phase
transition to become stable instead of $r_{+}=0.80.$ Therefore, we can say
that the quantum gravity effect grows the outer horizon $r_{+}$.

\section{Concluding remarks}\label{conc}

In this study, we have investigated the quantum gravity effect on the
thermodynamical properties of the CR-BTZ black hole in the context of the
quantum mechanical tunneling process of the charged massive spin-0 scalar particle, spin-1/2 Dirac particle and spin-1 vector boson
particle. According to the Eqs.(\ref{MEqq5}), (\ref{MEqq13}) and (\ref{MT}), some important results can be summarized as follows:

\begin{itemize}
\item Our calculations show that the modified Hawking temperature
of the CR-BTZ black hole depend not only on the black hole properties, but
also on the mass, charge, energy and angular momentum of the tunneling particle in the presence of the quantum gravity effect.
\item It is worth to note that the modified Hawking temperatures calculated via tunneling of scalar, Dirac and vector boson
particles are completely differently from each other.
\item The modified Hawking temperatures are lower than the standard Hawking temperature.
\item In the absence of the quantum gravity effect, i.e. $\alpha $=$0$, the modified Hawking temperatures are reduced to the standard Hawking temperature.
\end{itemize}

In this study, also, we analyse the local stability of the CR-BTZ black hole by using the modified Hawking temperature of the charged massive scalar particle.
We show that the modified heat capacity of the black hole depends on both the properties of the black hole and the
tunneling scalar particle, as well (see Eq.(\ref{capacity1})). Some important results given as follows:

\begin{itemize}
\item In the absence of the quantum gravity effect, i.e. $\alpha $=$0$, the modified heat capacity reduces to the standard one.
\item The black hole undergoes a first type of phase transition in order to become stable in the absence of the quantum gravity effect (dashed lines in Fig. 2 and 3).
\item On the other hand, it may undergo both first and second-type of phase transitions in order to become stable in the presence of the quantum gravity effect. A possible first-type phase transition for special value of the $E, J, j, l, q, Q, M_{0}, \alpha$ parameters is figured out in Fig. 1. Moreover, as shown in Fig. 2 and 3 by solid lines for another special value of these parameters, the black hole may undergo second-type phase transition, also.
\item According to Eq.(\ref{capacity1}), the modified heat capacity depends on the angular momentum, $J$, charge, $Q$, radius of the outer horizon, $r_{+}$, of the black hole, and the mass, $M_{0}$, angular momentum, $j$, charge, $q$ and total energy, $\widetilde{E}$, of the scalar particle. Also, besides these parameters, the modified heat capacity depends on the cosmological radius, $l$, and $\alpha$ parameter. The cosmological radius, $l$, $\alpha$ parameter, the radius of the outer horizon, $r_{+}$, the angular momentum, $j$, and the mass, $M_{0}$ of the particle are positive constant. Also, according to quantum tunnelling process of the particle from the black hole, the energy of the particle, $\widetilde{E}$, must be positive, i.e. $E-j\Omega_{+}>0$ \cite{20,23}. The only $J$, $Q$ and $q$ may take negative or positive value. Hence, according to choose of the special value of these parameter, we can say that the black hole will be undergo a first or a second-type of phase transition to become stable in the presence of the quantum gravity effect.
\item On the other hand, according to Eq.(\ref{capacity2}), in the absence of the quantum gravity effect, the positive or negative values of the $J$, $Q$ and $q$ do not affect the stability condition of the black hole.
\item From the solid lines in Fig. 2 and 3, we see that the quantum gravity effect grows the outer horizon $r_{+}$, i.e.,
it gives more deep insight about the black hole.
\end{itemize}

Finally, we can say that, thanks to the quantum gravity effect, the nature
of the tunneling particles play an important role in understanding the
evolution of a black hole such that it can shed some light on the final stage of
black holes.

\section*{Acknowledgements}

The authors are grateful to the anonymous referees for his/her comments and positive contribution to the context of the paper. This work was supported by Akdeniz University Scientific Research Projects
Unit and the Scientific and Technological Research Council of Turkey (TUBITAK) under Project Number 116F329.

\section*{Appendix}

\section*{A Tunneling of the charged massive Dirac particle from the CR-BTZ black hole}

In order to make a discussion on a tunneling Dirac particle from the CR-BTZ black hole under the GUP
effect, by means of the GUP relations, the standard Dirac equation \cite{57}
is be modified as follows;
\begin{eqnarray*}
-i\overline{\sigma }^{0}(x)\partial _{0}\widetilde{\Psi} =\left(i\overline{\sigma }^{i}(x)\partial _{i}-i\overline{\sigma}^{\mu}(x)\Gamma_{\mu }-\frac{m_{0}}{\hbar}-\overline{\sigma}^{\mu}(x)\frac{q}{\hbar}A_{\mu}\right) \left(1+\alpha\hbar^{2}\partial _{j}\partial ^{j}-\alpha m_{0}^{2}\right)\widetilde{\Psi}. \label{MDE1}
\end{eqnarray*}
Using the procedure that applied in \cite{50,51,52}, in the presence of the quantum gravity effect, the Hawking temperature of CR-BTZ black hole is
obtained as
\begin{eqnarray*}
T_{H}^{D}=\frac{T_{H}}{\left[1+\alpha \Pi\right]},
\end{eqnarray*}
or
\begin{equation}
T_{H}^{D}\simeq T_{H}\left[ 1-\alpha \Pi\right]. \label{MEqq13}
\end{equation}
where $T_{H}$ is the standard Hawking temperature given in Eq.(\ref{std}) and the $\Pi$ is
\begin{eqnarray*}
\Pi=\frac{3H(E-j\Omega_{+})m_{0}^{2}r_{+}^{2}-\left[(E-j\Omega_{+})(j^{2}H+4q^{2}Q^{2}r_{+}\ln(\frac{r_{+}}{l}))+4(E-j\Omega_{+})^{2}qQr_{+}\right]}{2r_{+}^{2}H\left(E-j\Omega_{+}+qQ\ln(\frac{r_{+}}{l})\right)}.
\end{eqnarray*}
This result shows that the modified Hawking temperature of the charged
massive Dirac particle is lower than the standard Hawking temperature, and
depends on both the black hole and the particle properties. Also, it is different from that of the scalar particle.

\section*{B Tunneling of the charged massive vector boson from the CR-BTZ black hole}

In the presence of the electromagnetic interaction, the standard massive vector boson equation given in
\cite{58} can be modified in the context of GUP relations as follows;
\begin{eqnarray*}
-i{\beta}^{0}(x)\partial_{0}\widetilde{\Psi}=\left(i\beta^{i}(x)\partial_{i}-i\beta^{\mu}(x)\Sigma_{\mu }-\frac{\mu_{0}}{\hbar}-\beta^{\mu}(x)\frac{q}{\hbar}A_{\mu}\right) \left(1+\alpha\hbar^{2}\partial_{j}\partial ^{j}-\alpha \mu_{0}^{2}\right)\widetilde{\Psi} \label{MSE1}
\end{eqnarray*}
Thus, using the procedure that applied in \cite{49}, the modified Hawking temperature of the CR-BTZ black hole obtained as
\begin{eqnarray*}
T_{H}^{^{VB}}=\frac{T_{H}}{\left[1+\alpha \Delta \right]},
\end{eqnarray*}
or
\begin{eqnarray}
T_{H}^{^{VB}}\simeq T_{H}\left[ 1-\alpha \Delta \right], \label{MT}
\end{eqnarray}
where $T_{H}$ is the standard Hawking temperature given in Eq.(\ref{std}) and $\Delta$ is
\begin{eqnarray*}
\Delta=\frac{(E-j\Omega_{+})\left[9r_{+}^{2}\mu_{0}^{2}H-4\left(4qQ(E-j\Omega_{+})r_{+}+j^{2}H+4r_{+}q^{2}Q^{2}\ln(\frac{r_{+}}{l})\right)\right]}{8r_{+}^{2}H\left(E-j\Omega_{+}+qQ\ln(\frac{r_{+}}{l})\right)}.
\end{eqnarray*}
This result shows that the modified Hawking temperature of the black hole obtained by the tunnelling process of the charged massive
vector boson particle is lower than the standard temperature and different
from that of the modified Hawking temperatures of the scalar and Dirac
particle.

\section*{C Explicit forms of the constants in the Eq.(\ref{capacity1})}
The constants in Eq.(\ref{capacity1}) are given as follows;
\begin{eqnarray*}
\mathcal{X} &=&3\alpha \widetilde{E}^{2}r_{+}^{2}l^{2}\left[J^{2}M_{0}^{2}+\frac{2}{3\alpha }\left(r_{+}^{2}Q^{2}+J^{2}\right)+\frac{J^{2}j^{2}}{r_{+}^{2}}%
+Q^{2}\left(j^{2}+M_{0}^{2}r_{+}^{2}\right)+\frac{4}{3}\widetilde{E}qQr_{+}^{2}\right]\\
&&+2r_{+}^{2}l^{2}qQ\ln(\frac{r_{+}}{l})\left\{2J^{2}\widetilde{E}\left[1+\frac{r_{+}^{2}Q}{J^{2}}\left(Q+\alpha \widetilde{E}qQ\right)+2\alpha j^{2}\left(\frac{1}{r_{+}^{2}}+Q^{2}\right)\right] \right. \\
&&\left. +qQ\ln(\frac{r_{+}}{l})\left[J^{2}\left(1+2\alpha j^{2}\left(\frac{1}{%
r_{+}^{2}}+Q^{2}\right)\right)+r_{+}^{2}Q^{2}\right] \right\}
\end{eqnarray*}
\begin{eqnarray*}
\mathcal{Y}=4r_{+}^{4}qQ\ln(\frac{r_{+}}{l})\left[\widetilde{E}\left(4r_{+}^{2}+8\alpha j^{2}\right)+2qQ\ln(\frac{r_{+}}{l})\left(r_{+}^{2}+2\alpha j^{2}\right)\right] +4r_{+}^{4}\widetilde{E}^{2}\left[r_{+}^{2}\left(2+3\alpha M_{0}^{2}\right)+3\alpha j^{2}\right],
\end{eqnarray*}
\begin{eqnarray*}
\mathcal{A}&=&r_{+}^{4}\widetilde{E}^{3}\left[
16r_{+}^{2}\left(2r_{+}^{4}+l^{2}J^{2}+9\alpha j^{2}r_{+}^{2}\right)+\alpha
Q^{4}l^{4}\left(3j^{2}+4q^{2}r_{+}^{2}+6qQM_{0}^{2}\widetilde{E}^{-1}+\frac{%
4q^{2}J^{2}}{Q^{2}}\right)\right]\\
&&+6\alpha\widetilde{E}^{2}J^{2}l^{2}M_{0}^{2}r_{+}^{2}\left[\left(4\widetilde{E}+\frac{%
8qQ^{3}j^{2}}{3J^{2}M_{0}^{2}}\right)r_{+}^{4}+qQ\left(J^{2}l^{2}+(\frac{8j^{2}}{%
3M_{0}^{2}}+2l^{2}Q^{2})r_{+}^{2}\right)\right]\\
&&+48r_{+}^{6}\widetilde{E}q^{2}Q^{2}\left(\ln (\frac{r_{+}}{l})\right)^{2}\left[
J^{2}l^{2}+\alpha j^{2}\left(12r_{+}^{2}+\frac{Q^{4}l^{4}}{4r_{+}^{2}}\right)+2r_{+}^{4}%
\right] \\ &&+4q^{3}Q^{3}r_{+}^{4}\left(\ln (\frac{r_{+}}{l})\right)^{3}\left[
8r_{+}^{6}+\alpha j^{2}\left(Q^{4}l^{4}+48r_{+}^{4}\right)+4J^{2}l^{2}r_{+}^{2}\right] \\
&&+12\alpha\widetilde{E}^{2}qQr_{+}^{6}\ln(\frac{r_{+}}{l})\left[\frac{1}{3}q^{2}Q^{4}l^{4}+4r_{+}^{4}\left(M_{0}^{2}+2\right)+44r_{+}^{2}j^{2}\right. \\
&&\left.+\frac{4}{\alpha }\left(J^{2}l^{2}+2r_{+}^{4}\right)+2J^{2}l^{2}M_{0}^{2}+\frac{%
Q^{3}l^{4}}{r_{+}^{2}}\left(\frac{11}{12}Qj^{2}+\frac{q^{2}}{3}J^{4}\right)\right]
\\ && +48\alpha \widetilde{E}^{2}M_{0}^{2}r_{+}^{10}\left(\widetilde{E}+2qQ\right),
\end{eqnarray*}
\begin{eqnarray*}
\mathcal{B}&=&16l^{2}q^{2}Q^{2}\widetilde{E}^{2}r_{+}^{6}\ln(\frac{r_{+}}{l})\left(11\alpha
q^{-1}Qj^{2}+\frac{3}{8}l^{2}q^{-1}Q^{3}+\alpha r_{+}^{2}\right)\left[ qQ+\widetilde{E}\left(6+\frac{Q^{2}l^{2}}{2r_{+}^{2}}\right)+\frac{3J^{2}l^{2}}{2\alpha
qr_{+}^{4}}\left(Q+\frac{3J^{2}}{4Qr_{+}^{2}}\right)\right]\\ &&+\alpha qQJ^{2}l^{4}\widetilde{E}^{2}\ln (\frac{r_{+}}{l})\left[M_{0}^{2}r_{+}^{2}\left(9J^{2}\widetilde{E}^{-1}+3Q^{2}r_{+}^{2}\left(4+Q^{2}J^{-2}r_{+}^{2}\right)\right)+40\widetilde{E}%
qQr_{+}^{4}+11\left(J^{2}j^{2}+8j^{2}l^{-2}r_{+}^{4}\right)\right]\\
&&+6l^{4}r_{+}^{2}q^{2}Q^{2}\widetilde{E}\left(\ln(\frac{r_{+}}{l})\right)^{2}\left[
3J^{4}+4Q^{2}J^{2}r_{+}^{2}+\left(\frac{2}{3}\alpha q\widetilde{E}%
+Q\right)Q^{3}r_{+}^{4}+\left(8r_{+}^{2}l^{-2}+\frac{10}{3}J^{2}\right)\alpha \widetilde{E}%
qQr_{+}^{4}\right]\\
&&+16\alpha\widetilde{E}^{2}q^{2}Q^{2}l^{2}r_{+}^{8}\left(1+3\widetilde{E}\left(\frac{\widetilde{E}}{qQ}+\frac{j^{2}}{q^{2}r_{+}^{2}}\right)\right)+4\alpha J^{2}l^{2}\widetilde{E}^{2}r_{+}^{2}\left[\left(6\widetilde{E}r_{+}^{2}+Q^{3}l^{2}\right)j^{2}+12qQM_{0}^{2}r_{+}^{4}\right]\\
&&+4\alpha q^{3}Q^{3}l^{4}\left(\ln (\frac{r_{+}}{l})\right)^{3}\left[J^{4}\left(j^{2}+%
\frac{3}{2\alpha }r_{+}^{2}\right)+Q^{2}r_{+}^{6}\left(16l^{-2}j^{2}+\frac{1}{2}%
Q^{2}\right)+J^{2}r_{+}^{4}\left(\frac{2}{\alpha }Q^{2}+8j^{2}l^{-2}\right)\right]\\
&&+Q^{4}l^{4}\widetilde{E}^{3}r_{+}^{6}\left[2r+3\alpha\left(M_{0}^{2}+\frac{J^{4}}{Q^{4}}\left(j^{2}r_{+}^{-2}+3M_{0}^{2}+6\right)r_{+}^{-4}\right)+8J^{2}Q^{-2}r_{+}^{-2}+4\alpha q\widetilde{E}Q^{-1}\right]\\
&&+2\alpha qJ^{2}l^{4}\widetilde{E}^{2}Q^{2}r_{+}^{4}\left[
\frac{6\widetilde{E}M_{0}^{2}}{q}+\frac{10\widetilde{E}^{2}}{Q}+\frac{j^{2}J^{2}}{Qr_{+}^{4}}+\frac{Q^{3}j^{2}}{J^{2}}+\frac{8Qr_{+}^{4}}{J^{2}l^{2}}\left(3M_{0}^{2}+2Q^{-2}j^{2}l^{-2}\right)\right]\\
&&+16\alpha q^{2}Q^{2}l^{2}\widetilde{E}\left(\ln (\frac{r_{+}}{l})\right)^{2}\left[
j^{2}J^{2}\left(6r_{+}^{4}+\frac{3}{4}J^{2}l^{2}\right)+\left(12r_{+}^{2}Qj^{2}+\frac{5}{4}%
\widetilde{E}J^{2}l^{2}q\right)r_{+}^{4}Q\right]\\.
\end{eqnarray*}
with $\widetilde{E}$=$E-j\Omega_{+}>0$ \cite{20,23}.

\end{document}